\documentclass[a4paper,twocolumn,prb,10pt]{{revtex4}}%,twocolumn
\usepackage{amsmath}
\usepackage{amssymb}
\usepackage{color}
\usepackage[utf8]{inputenc}
\usepackage{graphicx}

\usepackage{textcomp}

\begin{document}
\title{\sffamily 
Zener tunneling in the electrical transport of quasi-metallic carbon nanotubes}
\author{Gaston Kan\'{e}}
\affiliation{IMPMC, Universit\'{e} Pierre et Marie Curie-Paris 6,  CNRS, 4 place Jussieu, F-75005 Paris, France}
\author{Michele Lazzeri}
\affiliation{IMPMC, Universit\'{e} Pierre et Marie Curie-Paris 6,  CNRS, 4 place Jussieu, F-75005 Paris, France}
\author{Francesco Mauri}
\affiliation{IMPMC, Universit\'{e} Pierre et Marie Curie-Paris 6,  CNRS, 4 place Jussieu, F-75005 Paris, France}

\begin{abstract}

We study theoretically the impact of Zener tunneling 
on the charge-transport properties of
quasi-metallic (Qm) carbon nanotubes (characterized by forbidden band gaps of few tens of meV).
We also analyze the interplay between Zener tunneling and elastic scattering on defects.
To this purpose we use a model based on the master equation for the density matrix, that takes into account
the inter-band Zener transitions induced by the electric field (a quantum mechanical effect), the electron-defect scattering 
and the electron-phonon scattering.
In presence of Zener tunnelling the Qm tubes support an electrical current even when 
the Fermi energy lies in the forbidden band gap. In absence of elastic scattering (in high quality samples), the small size of the band gap of Qm tubes enables Zener tunnelling for realistic values of the the electric field (above $\sim$ 1 V/$\mu$m).  
The presence of a strong elastic scattering (in low quality samples) further decreases the values of the field required to observe Zener tunnelling. 
Indeed, for elastic-scattering lengths of the order of 50 nm, Zener tunnelling affects the current/voltage characteristic already in the linear regime.  In other words, in quasi-metallic tubes, Zener tunneling is made more visible by defects.
\end{abstract}	
\maketitle
%\begin{multicols}{2}

\section{Introduction}
Single-wall carbon nanotubes (SWNTs) are quasi one-dimensional wires
of great interest for future electronic-devices applications. 
A SWNT can be constructed by rolling up a graphene
sheet and its geometry is univocally specified by a pair of chiral indexes $(n,m)$~\cite{Roche}. 
The relationship between $n$ and $m$ defines three groups of SWNTS:
armchair $(n,m=n)$, zigzag ($n=0 $ or $m=0$) and chiral nanotubes.  
The indexes $(n,m)$ also determine the SWNT electronic structure. 
If $m-n$ is not a multiple of 3, SWNTs of diameters in the 1-2 nm range are 
semiconductors with a band gap $\Delta$ larger than 0.5 eV. 
Only armchair SWNTs were believed to be truly metallic.
Indeed, because of the curvature, zigzag and chiral SWNTs with $m-n$ multiple of 3,
present a small band gap of few tens of meV \cite{Min}, and for this reason are called
quasi-metallic (Qm)~\cite{Delaney}. 
However, recent transport measurements on 
chirality-identified tubes showed that also armchair tubes are Qm with $\Delta \sim 10$
to $100$~meV, likely because of electrons correlation \cite{Vikram}.

The description and interpretation of the transport properties of SWNT devices, are often
based on a semi-classical approach such as the Boltzmann transport
equation (BTE)~\cite{kane}.  However, the BTE can not describe  some cases  such us strong localization regimes~\cite{Roche}. Another effect that can not be described by the Boltzmann transport approach
is the Zener tunneling (ZT) \cite{Zener}, which is the electric-field-induced tunneling of carriers
from one band to another through the forbidden energy gap 
$\Delta$. In absence of ZT, if the temperature is smaller than the gap 
($ k_B T \ll \Delta$), only one type of carrier is responsible for charge transport:
when the Fermi level, $\epsilon_F$, is in the conduction (valence) band, the current is carried 
 by electrons (holes).
If $\epsilon_F$ is in the gap region charge-transport does not
occur. On the contrary, in presence
of ZT, the current is carried simultaneously by both electrons 
and holes and a non-zero
current exists even if $\epsilon_F$ is in the gap region. 

The critical electric field $E_z$ for which the ZT affects transport properties,
can be estimated considering the inter-band tunneling probability across the gap 
of the SWNT hyperbolic bands.
In absence of scattering, the transition probability as a function of the source-drain electric field $E_{SD}$
is given by
$T_{z}=\exp(-\pi E_z/E_{SD})$ \cite{Andreev, Jena} where 
\begin{equation}
E_z = \frac{\Delta^2}{ 4\hbar v_F e}
\label{Ezener}
\end{equation}
and $v_F\sim 10^8$~cm.$\mathrm{s}^{-1}$ is the
Fermi velocity of graphene (and of Qm nanotubes far from the gap). 
In semiconducting
tubes ($\Delta \sim 500$~meV),  $E_z \sim 95$~V/$\mathrm{\mu
m}$ is unrealistically large. At experimentally-accessible electric fields, ZT 
is absent, and in a diffusive regime
a semi-classical picture is sufficient to describe
transport properties. Instead, in Qm tubes ($\Delta \sim 50$~meV) $E_z \sim
0.95$~V/$\mathrm{\mu m}$, a value that nanotubes can sustain~\cite{kane}.
In this case ZT could be observed.

In this work, we study theoretically the current voltage characteristics, of Qm nanotubes both in the 
linear and high field regime and we analyze the interplay between Zener 
tunneling and defect (or impurity) scattering. 
We consider long Qm tubes with an energy band gap
$\Delta = 60$ $\mathrm{meV}$ (similar to the values measured in
experiments \cite{Min, Vikram}) in a diffusive regime, {\sl i.e.} when
the tube length is much longer than the defect-induced elastic scattering lengths ($l_e$) of 50~nm and 300~nm
presently used.
The approach is based on the solution of the master equation for the evolution of the 
density matrix. This approach correctly describes inter-band Zener tunneling.

\section{General assumptions}
\label{assump}
We take into account the elastic scattering of electrons with defects (we consider scattering  of electrons by short-range potential) and the inelastic scattering with optical phonons. 
We do not consider the possible occurence of weak or strong localization effects \cite{Roche}. Elastic and phonon scattering are assumed to be incoherent and, thus, the phase coherence length $L_\phi \leq l_e$.
Scattering with acoustic phonons plays a minor role since it is
characterized by scattering lengths that are much larger than the $l_e$
used here (see {\it e.g.} Ref.~\onlinecite{bushmaker09}). 
We suppose that the tubes are supported and that, for the regime under study, the phonon population 
is equilibrated at the temperature of the substrate.
We assume that the doping and the potential drop is uniform along the channel, and that the electron 
population of the valence and conduction bands
is independent on the position in the channel.
Namely, we consider the bulk transport properties of an infinite SWNT
under a spatially homogeneous electric-field applied along the tube axis. 

We remind that at high bias conditions in small gap nanotubes, the
scattering with optical phonons can lead to a current saturation~\cite{kane} and
to nonequilibrium optical-phonon population (hot phonons)~\cite{Lazzeri1,Lazzeri2}.
In suspended metallic tubes, this can lead to the observation of
negative differential resistance~\cite{pop05,bushmaker09}.
This kind of phenomena are not included in the present model and
their inclusion would lead to increase of the differential resistance
for high electric field (this point will be discussed in Sects.~\ref{sec_transport}-~\ref{sec_current}).

Electronic transport measurements in Qm nanotubes
were reported in ~\onlinecite{Vikram,bushmaker09}.
The two papers deal with transport regimes which are quite differents from
the present one. Because of this, the results of ~\onlinecite{Vikram,bushmaker09} can not be
directly compared with the present ones.
In particular, in both ~\onlinecite{Vikram,bushmaker09},
the elastic scattering lengths are much longer than in the present
work and, in both cases, they are comparable with the length of the
tubes. Because of this, Ref.~\onlinecite{Vikram} can observe interference effects
between the contacts and Ref.~\onlinecite{bushmaker09} can interpret the results
by assuming an electronic distribution non uniform along the
length of the channel. On the contrary, in the present work we are in a diffusive regime:
the scattering length $l_e\ll L$, where $L$ is the length of the tube,
and $L_\phi \ll L$, where $L_\phi$ is the phase coherence length.
In this diffusive regime, we can assume a uniform distribution of
the electrons along the tube and that interference effects between the contacts
are not observable. 

\section{Density matrix in an uniform electric field} 
\subsection{ Hamiltonian } 

The graphene Hamiltonian for
the $\pi$-bands close to the $\mathbf K$ point (Dirac point) of the
Brillouin zone (BZ) is \cite{Castro}:
\begin{equation} \mathcal{H}_{2D} = v_F\hat{\sigma}\cdot \bf p
\label{graph}
\end{equation} where $v_F$ is the Fermi velocity,
$\hat{\sigma}=(\sigma_x,\sigma_y)$ with $\sigma_x$ and $\sigma_y$ the
Pauli matrices. ${\bf p}$ is the momentum operator in the
two-dimensional BZ of graphene. $\mathcal{H}_{2D}$ is a 2$\times$2
matrix which describes the $\pi$ and $\pi^*$ bands. Starting from
Eq.(\ref{graph}), one can derive an effective Hamiltonian
$\mathcal{H}$ for a SWNT:
\begin{equation} \mathcal{H} = v_F (\mathbf{\sigma}_x p_x +
\mathbf{\sigma}_y \hbar k_o )
\label{nanot}
\end{equation} $p_x$ is the momentum operator in the one-dimensional
BZ of the nanotube (a momentum parallel to the tube axis) and $k_o$ is
a constant fixed by the energy Gap $\Delta =2\hbar v_F k_o$. The
eigenstates of $\mathcal{H}$ are Bloch states and have the form
\begin{equation} \vert k, \alpha\rangle = \frac{ e^{i k x}}{\sqrt{L}}
\vert U_{k, \alpha}\rangle.
\end{equation} 
Here $L$ is the length of the system and $\vert U_{k,
\alpha}\rangle$, is the periodic part of the Bloch state:
\begin{equation} \vert U_{k, \alpha}\rangle = \frac{1}{\sqrt{2}}
\left( \begin{array}{c} 1 \\ \alpha e^{-i\theta}
\end{array} \right)
\label{period}
\end{equation} where $\theta =\arctan(k_o/k)$. $\alpha = 1$ indicates
positive energy (conduction band) whereas $ \alpha = -1$ describes
states with negative energy (valence band) with respect to the Dirac
point.  The eigenvalues of $\mathcal{H}$ read:
\begin{equation} \epsilon_{\alpha} (k)=\alpha \hbar v_F \sqrt{k ^2 +
k^2_o }
\label{nanote}
\end{equation} We now apply a spatially homogeneous and static
(source-drain) electric field $E_{SD}$ along the tube axis. The Hamiltonian, in the
vector-potential gauge, becomes :
\begin{equation} \overline{\mathcal{H}} = v_F [\mathbf{\sigma}_x (p_x+
e\mathcal{A} (t)/c) + \mathbf{\sigma}_y \hbar k_o ]
\label{nanotfield}
\end{equation} Where $\mathcal{A}(t)= -c E_{SD} t$ is the vector potential
component along the tube axis, and $e=\vert e\vert$ the electron
charge. The instantaneous eigenstates of the time dependent
Hamiltonian, Eq.(\ref{nanotfield}), are given by \cite{krieger1} :
\begin{equation} \overline{\vert k, \alpha \rangle}=
e^{-ie\mathcal{A}(t)x/\hbar c}\vert k, \alpha\rangle
\label{nanotfb}
\end{equation} with eigenvalues $\epsilon_{\alpha} (k)$.
\subsection{High field: non-linear regime} 
A framework in which quantum effects
such as Zener tunneling appear naturally is the Liouville-Von Neumann
Master Equation (ME) \cite{Rossi}. Within this formalism our variable
is the one-electron density-matrix operator $\rho$. 
The time evolution of $\rho$
in absence of scattering
processes is given by:
\begin{equation} \frac{\rm{d}\rho}{\rm{d}t} =
-\frac{i}{\hbar}[\overline{\mathcal{H}} ,\rho].
\label{liouville}
\end{equation} 
Since we are
dealing with a spatially homogeneous electric field applied to an
infinite nanotube, only the elements of the density
matrix diagonal with respect to $k$ will be relevant \cite{Rossi2}.
We use the instantaneous eigenstates (Eq.(\ref{nanotfb})) as
basis. Then, from Eq.(\ref{liouville}) one finds the time evolution of
$\rho_{\alpha\beta}(k, t) =
\overline{\langle k,\alpha \vert} \rho \overline{ \vert k,
\beta\rangle}$ \cite{krieger2}:
\begin{equation}
\begin{split} \frac{\partial {\rho }_{\alpha \beta }(k,t) }{\partial
t}+ \frac{eE_{SD}}{\hbar }\frac{\partial \rho _{\alpha \beta
}(k,t)}{\partial k}=\frac{\Delta \epsilon _{\alpha \beta }}{i\hbar
}\rho _{\alpha \beta }(k,t) & \\+\frac{eE_{SD}}{i\hbar }\sum\limits_{\eta
}(R_{\alpha \eta}(k)\rho_{\eta\beta}(k,t)-\rho_{\alpha\eta
}(k,t)R_{\eta \beta }(k))
\end{split}
\label{me1}
\end{equation} where $\Delta \epsilon_{\alpha \beta} =
\epsilon_{\alpha} (k) -\epsilon_{\beta} (k)$ and $R_{\alpha \eta }$ is
defined by:
\begin{equation} R_{\alpha \eta}(k) = \langle U_{k,\alpha}\vert
\frac{1}{i} \frac{\partial }{\partial k} \vert U_{k, \eta}\rangle
(1-\delta_{\alpha \eta}) .
 \label{Rzen1}
\end{equation} 
The $R_{\alpha \eta }$ term introduces the inter-band
transitions.  
Using perturbation theory for the derivative with
respect to $k$ we can show that:
\begin{equation} R_{\alpha \eta }(k) = i\hbar\frac{\langle k,\alpha
\vert v \vert k, \eta\rangle}{\epsilon_{\alpha} (k) -\epsilon_{\eta}
(k)}(1-\delta_{\alpha \eta})
\label{Rzen}
\end{equation} where $v$ is the velocity operator defined by:
\begin{equation} v =\frac{i}{\hbar}[\mathcal{H}, x] =v_F\sigma_x
\label{velop}
\end{equation} 
For our system $R_{\alpha \eta }$ takes the form:
\begin{equation} R_{\alpha \eta }(k)=-\frac{1}{2}\frac{k_o}{k^2 +
k^2_o} (1-\delta_{\alpha \eta}).
\end{equation} 

The full evolution equation is obtained by adding the
collision integral $ I[\rho]_{\alpha\beta}$ to the right-hand side of
Eq.(\ref{me1}). The collision integral results from the sum of two
independent contributions due to inelastic electron-phonon scattering
and elastic electron-defect scattering. Electron-phonon scattering is
treated as in Ref. \cite{graham} and read:
\begin{equation}
\begin{split} \mathrm{I}[\rho]^{e-ph}_{\alpha\beta}=\frac{1}{2L} \sum_
{k'} \sum_{\eta \nu}(W^{\nu}_{k'\eta \rightarrow k\alpha }
+W^{\nu}_{k'\eta\rightarrow k\beta })& \\ \times \rho _{\eta \eta
}(k')(\delta _{\alpha\beta }-\rho_{\alpha \beta }(k)) &
\\-\frac{1}{2L} \sum_{k'}\sum_{\eta \nu}(W^{\nu}_{k\alpha\rightarrow
k'\eta } +W^{\nu}_{k\beta\rightarrow k'\eta })& \\ \times \rho
_{\alpha\beta }(k)(1-\rho _{\eta\eta}(k'))
\end{split}
\label{colph}
\end{equation} where $W^{\nu}_{k\alpha\rightarrow k'\eta }$, the
electron-phonon scattering probability is evaluated using Fermi golden
rule \cite{Hamaguchi}:
\begin{equation}
\begin{split} W^{\nu}_{k'\eta \rightarrow k\alpha}=\frac{2\pi}{\hbar}
\rvert g^{\nu}_{\eta\alpha}(k', k) \lvert^{2} \left(n_{q}^{\nu} +1/2
\pm 1/2 \right) & \\ \times
\delta(\epsilon_{\alpha}(k)-\epsilon_{\eta}(k')\pm \hbar
\omega_{q}^{\nu})
\end{split}
\label{probph}
\end{equation} with $g^{\nu}_{\eta\alpha}$ the coupling matrix
elements and $\nu$ the phonon branch. $\hbar \omega_{q}^{\nu}$ is the
phonon energy with $q=\rvert k' - k\lvert$ and $n_{q}^{\nu}$ is the
phonon population.  The upper (lower) sign in the delta function
corresponds to phonon emission (absorption). \\

The elastic scattering by defects is supposed to be
incoherent. In this case the collision integrals for the diagonal and
non-diagonal elements $(\alpha \neq \beta)$ of the density matrix are
given by:
\begin{equation} \mathrm{I}[\rho]_{\alpha \alpha} = \frac{1}{L}
\sum_{k'} W^{el}_{k'\alpha \rightarrow k\alpha} [ \rho_{\alpha
\alpha}(k') - \rho_{\alpha \alpha}(k)]
\end{equation}
\begin{equation} \mathrm{I}[\rho]_{\alpha\beta} = -\frac{1}{2L}
\sum_{k'}(W^{el}_{k'\alpha \rightarrow k\alpha } + W^{el}_{k'\beta
\rightarrow k\beta})\rho_{\alpha \beta}(k).
\end{equation} 
The transition
probability, within the Fermi golden rule, is defined by :
\begin{equation} 
W^{el}_{k'\alpha \rightarrow k\alpha } = \frac{2\pi
L^2 N_D}{\hbar}\rvert \langle k',\alpha|V_D|k,\alpha\rangle
\lvert^{2}\delta(\epsilon_{\alpha}(k')-\epsilon_{\alpha}(k))
\label{probel}
\end{equation} 
Where $N_D$ is the density of defects (that is the
number of defects per unit length) and $\rm{V_D}$ is the scattering
potential.

Finally, the master equation, that describes the time evolution 
reads:
\begin{equation}
\begin{split} \frac{\partial {\rho }_{\alpha \beta }(k) }{\partial t}+
\frac{eE_{SD}}{\hbar }\frac{\partial \rho _{\alpha \beta }(k)}{\partial
k}=\frac{\Delta \epsilon _{\alpha \beta }}{i\hbar }\rho _{\alpha \beta
}(k) & \\+\frac{eE_{SD}}{i\hbar }\sum\limits_{\eta }(R_{\alpha
\eta}(k)\rho_{\eta\beta}(k)- & \\ \rho_{\alpha\eta }(k)R_{\eta \beta
}(k)) +\mathrm{I}[\rho]^{el}_{\alpha \beta} +
\mathrm{I}[\rho]^{e-ph}_{\alpha\beta}.
\end{split}
\label{me2}
\end{equation} 
For a given value of the
electric field $E_{SD}$, a steady-state solution of Eq.(\ref{me2}) is reached
for $t\rightarrow\infty$.  In this work, the steady-solution 
is obtained by numerical
integration in time using a finite-difference method.

The electronic population (that is the diagonal
elements $\rho_{\alpha \alpha}$) determines the density of electrons
$N$ (number of electrons in excess per unit length of the tube) by:
\begin{equation} N = \frac{g_s g_v}{L}
\sum_{k}\left[\sum_{\alpha}\rho_{\alpha\alpha}(k)-1\right],
\end{equation} where $g_s =2$ and $g_v=2$ are respectively the spin
and valley degeneracy. $N$ is a constant of motion of Eq.(\ref{me2}).
We will show simulations done at different values of the
Fermi level $\epsilon_F$, that is by fixing the initial value of $N$
to:
\begin{equation} N=\frac{g_s g_v}{L}
\sum_{k}\left[\sum_{\alpha}f(\epsilon_{\alpha}(k)-\epsilon_F)-1
\right]
\end{equation} Where $f(\epsilon)$ is the Fermi-Dirac function, computed here for $T=50$~K.

The charge current is given by :
\begin{equation} I= \frac{-g_s g_v e}{L}
\sum_{k}\sum_{\alpha\beta}v_{\alpha\beta}(k)\rho_{\alpha\beta }(k),
 \label{curr}
\end{equation} where $v_{\alpha\beta}$ is defined by :
\begin{equation} v_{\alpha\beta}(k) = \overline{\langle k,\alpha
\vert} \frac{i}{\hbar}[\overline{ \mathcal{H}}, x] \overline{ \vert k,
\beta\rangle} =\langle k,\alpha \vert \frac{i}{\hbar}[\mathcal{H},x]
\vert k, \beta\rangle
\end{equation} 
Using the definition of the velocity operator given in
Eq.(\ref{velop}), one can show that
\begin{equation} \hbar v_{\alpha\beta}(k) = \left\{
\begin{array}{ll}
 \partial{\epsilon^o_{\alpha} (k)}/\partial{k} & \mbox{if } \alpha
=\beta \\ -i R_{\alpha \beta } (k)\Delta \epsilon_{\alpha \beta} &
\mbox{if } \alpha \neq \beta
\end{array} \right.
\label{veloc}
\end{equation} 
Finally the steady current will be evaluated for the
steady-state solution.

An interesting aspect of the master equation, is that we can recover the
 standard Boltzmann transport equation (BTE), i.e. the semiclassical 
description, setting $R_{\eta \beta }$ of Eq.(\ref{me2}) to zero.
In this case the diagonal elements of the density matrix evolve according to:
\begin{equation}
\begin{split} \frac{\partial {\rho }_{\alpha \alpha }(k) }{\partial
t}+\frac{eE_{SD}}{\hbar }\frac{\partial \rho _{\alpha \alpha
}(k)}{\partial k}=\mathrm{I}[\rho]^{el}_{\alpha \alpha} +
\mathrm{I}[\rho]^{e-ph}_{\alpha\alpha},
\end{split}
\label{boltz}
\end{equation} 
and the off-diagonal elements of the steady solution vanish.

\subsection{Linear regime} 

In this section, we derive a
set of simplified equations (which can be evaluated analytically)
valid for small values of the electric field $E_{SD}$ within the linear
regime.  In the linear regime the applied field will cause a small
deviation of the equilibrium distribution $(E_{SD}=0)$. The steady state
density matrix (with $E_{SD}\neq 0$) can be written as:
\begin{equation} {\rho }_{\alpha \beta}(k) = {\rho}^{(0)}_{\alpha
\beta}(k)+ {\rho}^{(1)}_{\alpha \beta}(k)
\label{eqlin1}
\end{equation} where ${\rho}^{(0)}_{\alpha \beta}$ is the density
matrix element in absence of field $(E_{SD}=0)$ and ${\rho}^{(1)}_{\alpha
\beta}$ the deviation linear in $E_{SD}$. ${\rho}^{(0)}_{\alpha \beta}$ is
the equilibrium distribution which is diagonal and equal to
${\rho}^{(0)}_{\alpha \beta} = \delta_{\alpha
\beta}f(\epsilon_{\alpha}(k)-\epsilon_F)$. Inserting Eq.(\ref{eqlin1})
in Eq.(\ref{me2}), and keeping all the terms linear in $E_{SD}$, one obtains
the solution for the diagonal elements of the density matrix:
\begin{equation} \rho^{(1)}_{\alpha \alpha}(k) =
\tau_r(k)\frac{eE_{SD}}{\hbar }\frac{\partial \rho^{(0)}
_{\alpha\alpha}(k)}{\partial k}
\label{diagdens}
\end{equation} with $\tau_r$, the momentum relaxation time, defined
by \cite{Hamaguchi}:
\begin{equation} \frac{1}{\tau_r(k)} = \frac{1}{L} \sum_{k'}
W^{el}_{k\alpha \rightarrow k'\alpha }\varphi(k,k')
 \label{relaxt}
\end{equation} where $\varphi(k,k') = 2$ for back-scattering processes
and $0$ for forward scattering ones.  Indeed, after the scattering
with a defect (or a phonon), the electron can either maintain its
propagation direction, which is forward scattering (fs), or reverse it,
which is back-scattering (bs).  The off-diagonal elements $(\alpha \neq
\beta)$ are given by:
\begin{equation} 
\rho^{(1)}_{\alpha\beta}(k) =\frac{eE_{SD} R_{\alpha
\beta}(k)}{\Delta\epsilon _{\alpha\beta} -
i\hbar\gamma_{tot} (\rho^{(0)} _{\alpha\alpha }(k)-\rho^{(0)}
_{\beta\beta }(k))}
\label{nondiagd}
\end{equation} where $\gamma_{tot}$ is defined by :
\begin{equation} \gamma_{tot}(k)=\gamma_{el}(k)+\gamma^{h}_{eph}
(k)+\gamma^{e}_{eph} (k)
\label{gamt}
\end{equation} where $\gamma_{el}$, $\gamma^{h}_{eph}$ and
$\gamma^{e}_{eph}$, are respectively the elastic, hole-phonon and
electron-phonon scattering rate for the equilibrium distribution and
defined by:
\begin{equation} \gamma_{el}(k) = \frac{1}{2L}
\sum_{k'}(W^{el}_{k\alpha \rightarrow k'\alpha} + W^{el}_{k\beta
\rightarrow k'\beta})
\label{gamel}
\end{equation}

\begin{equation} \gamma^{h}_{eph}(k) =\frac{1}{2L}
\sum_{k'}\sum_{\eta\nu}(W^{\nu}_{k'\eta \rightarrow k\alpha} +
W^{\nu}_{k'\eta \rightarrow k \beta})\rho^{(0)}_{\eta \eta}(k')
\label{gamephh}
\end{equation}

\begin{equation} \gamma^{e}_{eph}(k)=\frac{1}{2L}
\sum_{k'}\sum_{\eta\nu}(W^{\nu}_{k\alpha \rightarrow k' \eta}+
W^{\nu}_{k\beta \rightarrow k' \eta })[1-\rho^{(0)}_{\eta \eta}(k')]
\label{gamephe}
\end{equation}

\subsection{Zero-field electrical conductivity}
\label{sec_zero_field}
The zero-field conductivity, 
$\sigma^o=\lim_{E_{SD}\rightarrow 0} I/E_{SD}$, can  be obtained by inserting
the steady state solutions of the linear regime, 
Eqs.(\ref{diagdens}) and (\ref{nondiagd}) in the current expression,
Eq.(\ref{curr}):
\begin{equation} \sigma^o =\sigma_b ^{o}+\sigma_z ^{o}
\label{condtot}
\end{equation} where $\sigma_b ^{o}$ is the conductivity
 obtained from the diagonal elements of the density matrix.
It is given by:
\begin{equation} \sigma_b ^{o}= \frac{4 e^2}{L} \sum_{k}
\sum_{\alpha}v_{\alpha\alpha}(k)\frac{\tau_r(k)}{\hbar
}\left(-\frac{\partial \rho^{(0)} _{\alpha\alpha}(k)}{\partial
k}\right)
 \label{condbg}
\end{equation} $\sigma_b ^{o}$ is the semi-classical
conductivity of the Boltzmann equation and at zero temperature it reduces to:
\begin{equation} \sigma_b ^{o}= \frac{e^2
v_{11}^2(k_F)}{2}\tau_r(k_F)\mathrm{D}(\epsilon_F)
\label{condb}
\end{equation} with $\hbar k_F$ the Fermi momentum and
$\mathrm{D}(\epsilon)$ the one-dimensional density of states (DOS) per
unit length.  For the lowest sub-band of the Qm nanotube \cite{Lundstrom}:
\begin{equation} \mathrm{D}(\epsilon)
= \frac{\mathrm{D}_o \vert
\epsilon\vert}{\sqrt{\epsilon^2 - (\Delta/2)^2}}\Theta(\vert
\epsilon\vert - \Delta/2)
\label{Dos}
\end{equation} where $\mathrm{D}_o =\frac{4}{\pi \hbar v_F}$ is the
constant DOS (per unit length) of the lowest sub-band of a metallic
tube. $\Theta(x)$ is the step function which equals 1 for
$x>0$ and $0$ otherwise.  In the case of a metallic tube the
conductivity, Eq.(\ref{condb}), can be written as:
\begin{equation} \sigma_m ^{o}= \frac{2e^2}{\pi \hbar}l_e(k_F)
\label{sigm}
\end{equation} where $l_e(k_F)$, the elastic scattering length, is given
by:
\begin{equation} l_e(k_F) = v_F \tau_r (k_F).
\label{elastle}
\end{equation} 

The current due to Zener inter-band is given by $\sigma_z ^{o}$, that arises from
the off-diagonal part of the density matrix, Eq.~(\ref{nondiagd}):
\begin{eqnarray}
\sigma_z ^{o} &=& -\frac{4\hbar e^2}{L} \sum_{k}\sum
_{\alpha \beta} |\langle k,\alpha\vert v \vert k, \beta\rangle|^2\frac{\rho^{(0)}_{\alpha \alpha}(k)-\rho^{(0)} _{\beta
\beta}(k)}{\epsilon_{\alpha} (k) -\epsilon_{\beta} (k)} \nonumber\\
& &
\times
\frac{\hbar\gamma_{tot}(k)}
{[\epsilon_{\alpha} (k)
-\epsilon_{\beta} (k)]^2 + [\hbar\gamma_{tot}(k)]^2}(1-\delta_{\alpha \beta}).
\label{kubo}
\end{eqnarray} 
Notice that this expression
is similar to the off-diagonal Kubo formula for
the optical conductivity \cite{Nomura} but with a finite value of the
parameter $\gamma_{tot}$. When $\gamma_{tot} = 0$, the Zener contribution ($\sigma_z$)
vanishes and the total conductivity is equal to that of the semiclassical Boltzmann approach
($\sigma_b ^{o}$). 
Thus, in the linear regime, the Zener tunnelling contributes to the transport only
in presence of scattering events that limit the carrier lifetime. On the contrary, at finite field, Zener tunneling can occur in the ballistic case (see formula right above Eq.(\ref{Ezener})).

\section{Transport calculations}
\label{sec_transport} 

We performed calculations for
a metallic tube and a quasi-metallic tube (Qm) with an energy band gap
$\Delta = 60$ $\mathrm{meV}$ (similar to the values measured in
experiments \cite{Min, Vikram}). 
The tubes are subject to a spatially
uniform electric field and the
number of carriers (electrons or holes) is varied by changing the
Fermi level $\epsilon_F$.  In the metallic tube the dispersion
relation is linear $\epsilon_{\pm}(k)=\pm\hbar
v_Fk$, with $v_F=8.39
\times 10^7 $ $\mathrm{cm/s}$, in a Qm tube the bands are hyperbolas, Eq.(\ref{nanote}).

The transport properties of the Qm nanotube are
explored by a semi-classical and a quantum-mechanical approach. In the
semi-classical case, we do not take into account the inter-band
transitions due to the electric field by setting to zero the
off-diagonal elements of the density matrix ($\rho_{\alpha \beta}$
with $\alpha \neq \beta$ ). The evolution of the system is then governed by the BTE, Eq.~(\ref{boltz}).
In the quantum approach, we take into account the inter-band
transitions due to the electric field by considering the
off-diagonal elements of the density matrix. The
evolution is then governed by the ME, Eq.~(\ref{me2}).

\begin{figure}
\begin{center} \vspace*{0.9cm}
\includegraphics[width = 0.51\textwidth]{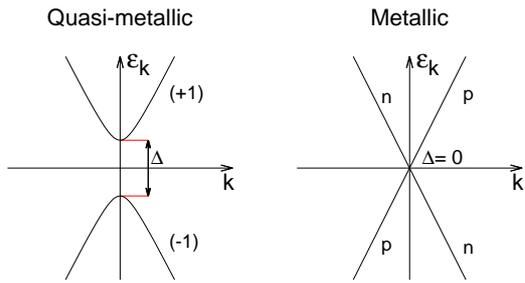}
\caption{\small Electronic band structures of a quasi-metallic
($\Delta \neq 0$) and a metallic carbon nanotube ($\Delta = 0$).  For
the Qm tube, we distinguish between valence and conduction bands
[labeled (-1) and (+1)]. On the contrary, for the metallic tube
it is more convenient to distinguish between bands in which the sign
of the electron velocity is the same, that is negative (n) or positive
(p).}
\label{qmb}
\end{center}
\end{figure}

For the metallic tube, the ME that governs the quantum evolution can be obtained in the
limit $\Delta\rightarrow 0$.
In this case,
the coupling $R_{\alpha \eta}$ [(Eq.~(\ref{Rzen1})], 
that promotes Zener transitions, diverges at $k=0$, and the electrons
that cross the $k=0$ point tunnel with 100\% probability from the valence (conduction) to the conduction (valence) band.
As a consequence, for 
metallic tubes, the quantum-mechanical approach (ME) reduces to the 
semiclassical approach (BTE) if we change the meaning of the band indexes in Eq.~\ref{boltz}.
In particular, while for Qm tubes we distinguish between ``valence''
and ``conduction'' bands, for metallic tubes we
distinguish between bands with positive or negative velocity 
($\partial\epsilon/\partial k < 0$ or $\partial\epsilon/\partial k > 0$), see Fig.\ref{qmb}.

We consider short-range scatterers and the electron-defect coupling 
$\rvert \langle k',\alpha|V_D|k,\alpha\rangle \lvert$
is taken to be a constant and $k$-independent. 
Under this hypothesis in a metallic tube the  elastic scattering mean free path
$l^{m}_e=\tau^{m}_r v_F$ is $k$-independent.
The value of $\rvert \langle k',\alpha|V_D|k,\alpha\rangle \lvert$ is determined fixing 
$l^{m}_e = 50$ nm and $l^{m}_e = 300$ nm.
In the Qm tubes, $\rvert \langle k',\alpha|V_D|k,\alpha\rangle \lvert$
is also assumed to be a constant equal to
that of the metallic tubes. 
Note, however, that, in the Qm case, the
scattering mean free path
become $k$-dependent. Indeed, because the presence of a gap, the
density of states, that determines the scattering rate, is energy dependent.

Only three optical phonons can play an important role in transport phenomena at high bias: 
two (longitudinal and transverse) correspond to the graphene optical phonons at
$\mathbf\Gamma$ with symmetry E$_{2g}$,
and one corresponds to the graphene phonon at $\mathbf{K}$ with
symmetry A$'_1$~\cite{kane, Euen, Lazzeri1, Lazzeri2}. The energies of
these phonons are respectively $\hbar\omega ^{\mathbf \Gamma}=
200$ $\mathrm{meV}$ and $\hbar\omega^{\mathbf K}=150$ $\mathrm{meV}$.
For our calculation the
electron-phonon coupling matrix elements $g^{\nu}_{\eta\alpha} (k',k)=
g^{\nu}_{A}$ are taken constant and independent of $k$ and
$k'$, with $A=bs,fs$. 
The label $bs$ stands for a back-scattering process 
($v_{\eta\eta}(k) v_{\alpha\alpha}(k')<0$), 
and $fs$ stands for a forward scattering process
($v_{\eta\eta}(k) v_{\alpha\alpha}(k')>0$).  
In a metallic tube, the
corresponding scattering lengths, $l^{\nu}_{A}$, are proportional to the tube diameter
$d$ \cite{Lazzeri2}, that we assume to be 2~nm.
In this case, $l^{\mathbf
\Gamma}_{bs}=l^{\mathbf \Gamma}_{fs} = 451.38$~ nm, $l^{\mathbf
K}_{bs}=183.74$~nm and $l^{\mathbf K}_{fs}=0$. For the Qm tubes we use the $g^{\nu}_{A}$ obtained in the metallic case.
The phonon population $n^\nu_q$ is set
to zero (cold phonons, only phonon emission is treated). We do not
consider the generation of hot phonons, that occurs at high bias in nanotubes, when the elastic scattering length is much larger than the phonon one, $l^{\nu}_{A}$ ~\cite{Lazzeri2}.
In the present case, this is justified by the strong elastic scattering used in
the present model that competes with the phonon scattering, as in graphene~\cite{Amelia}.  

\section{Results and discussion}

We now show the results obtained for two different values of the elastic
scattering length, namely $l_e = 50$~nm and $l_e = 300$~nm.
The goal is to understand in which conditions the Zener tunneling is
experimentally observable.
We will compare the results obtained with the
semi-classical (or Boltzmann) approach with those obtained with the
quantum treatment (master equation).
In the semi-classical approach the Zener tunneling is not allowed.
Thus, the semi-classical and quantum results will differ when Zener tunneling is relevant.
Results for metallic tubes (for which the distinction quantum versus. semi-classical does not hold)
are shown as a comparison.
The results shown for a finite value of the source-drain electric
field, $E_{SD}$, are obtained from numerical solution of the transport equations.
The results for $E_{SD}=0$ are derived analytically in Sec.~\ref{sec_zero_field}

\subsection{Conductivity and current as a function of the Fermi energy} 

\begin{figure}
\begin{center}
\includegraphics[width = 0.48\textwidth]{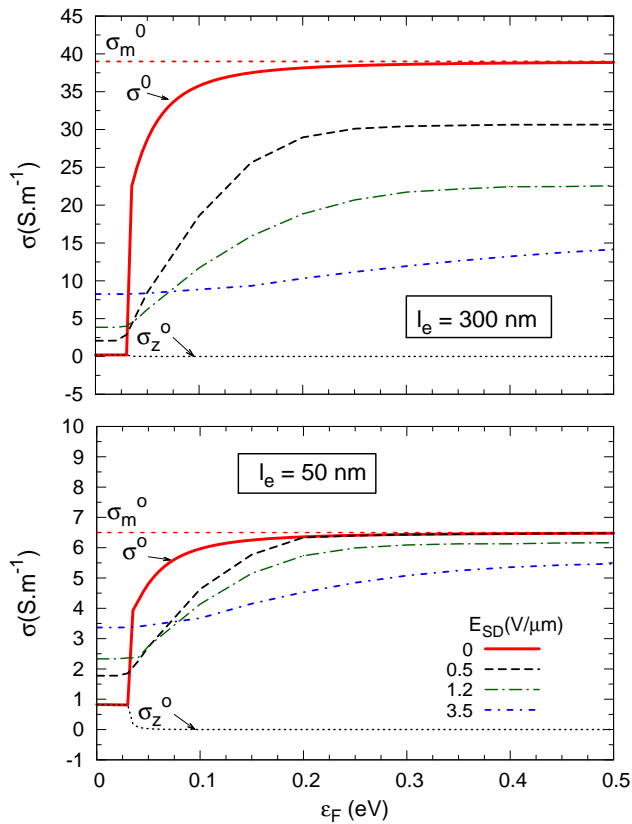}
%\vspace*{-0.52cm}
\caption{\small (Color online) Finite field conductivity
$\sigma=I/E_{SD}$ versus the Fermi energy $\epsilon _F$ for different values of the
field $E_{SD}$. The two panels report results obtained with two
different values of the elastic scattering length $l_e$. $\sigma_m^{0}$ and $\sigma^{0}$ are
the zero field conductivity of the metallic tube and of the Qm tube, respectively.
$\sigma_z^{0}$ is the component of $\sigma^0$ due to Zener tunneling.}
\label{cond5}
\end{center}
\end{figure}

%\begin{figure}
%\begin{center}
%\includegraphics[width = 0.48\textwidth]{fig2b.eps}
%\vspace*{-0.52cm}
%\caption{\small (Color online) Zero field conductivity 
%versus the Fermi energy $\epsilon _F$ . Results are obtained with two
%different values of the elastic scattering length $l_e$ in the region of 
%Fermi energy where ZT is important. $\sigma^{0}$ is
%the zero field conductivity of the Qm tube, and 
%$\sigma_z^{0}$ is the component of $\sigma^0$ due to Zener tunneling.
%The zero-bias conductivity for the smallest scattering length, $l_e=50$~nm
%is much higher than the one for the longest scattering length,
%$l_e=300$~nm.}
%\label{cond5b}
%\end{center}
%\end{figure}

Fig.~\ref{cond5} reports the conductivity $\sigma$ of the Qm tube as a function of
the Fermi level $\epsilon_F$ for various values of the electric field $E_{SD}$. 
In Fig.~\ref{cond5}, $\sigma^0$ is the zero-bias ($E_{SD}=0$) conductivity
calculated in the linear regime for the temperature $T=50$~K. 
$\sigma_z^0$ is the component of $\sigma^0$ due to Zener tunneling
($\sigma_z^0$ is a well defined quantity, see Sec.~\ref{sec_zero_field}),
and $\sigma_m^0$ is the zero-bias conductivity of a metallic tube.

For the metallic tube, the conductivity $\sigma_m^0$ has a finite
value which is proportional to the elastic scattering length $l_e$ and
which does not depend on $\epsilon_F$ (this is true as soon as $\epsilon_F$ does not meet the higher  subbands of the SWNT, which are not considered in this study).
For the Qm tube, the conductivity $\sigma^0$ tends to be equal to the
metallic one for $\epsilon_F \gg \Delta/2$: when the doping 
is sufficiently high there is only one type of carriers, the details
of the electronic bands near the Dirac point are not relevant,
the transport is properly described by the semi-classical 
Boltzmann  approach and the
Qm and metallic tube behavior are indistinguishable.
On the other hand, for $\epsilon_F \leq \Delta/2$ the presence of the
electronic gap $\Delta$ acts as a barrier and the Qm conductivity suddenly diminishes.
In particular, from Fig.~\ref{cond5}, 
for $\epsilon_F \leq \Delta/2$, $\sigma^0=\sigma_z^0$ meaning that the Qm
conductivity is entirely due to Zener tunneling (inter-band) and that
the Boltzmann conductivity (intra-band) vanishes.
Fig. \ref{cond5} also reports the finite field
conductivity $\sigma=I/E_{SD}$ for different values of the electric
field. Also at finite field, the conductivity is entirely due to Zener tunneling
for small values of $\epsilon_F$.

The comparison of the results obtained for different values of the
elastic scattering lengths, $l_e$, is very remarkable.
Indeed, for small doping ($\epsilon_F \leq \Delta/2$), the zero-bias conductivity
for the smallest scattering length
($\sigma^0 \simeq 0.82$~S.m$^{-1}$ for $l_e=50$~nm in Fig.~\ref{cond5}),
is higher than the one for the longest scattering length,
($\sigma^0 \simeq 0.2$~S.m$^{-1}$ for $l_e=300$~nm).
Moreover, most important, the ratio between the high doping $\sigma^0$ ($\epsilon_F>$0.2~eV)
and the small doping $\sigma^0$ is much higher for the smallest scattering length
than for the longest one.

\begin{figure}
\begin{center} %\vspace*{-0.cm}
\includegraphics[width = 0.44\textwidth]{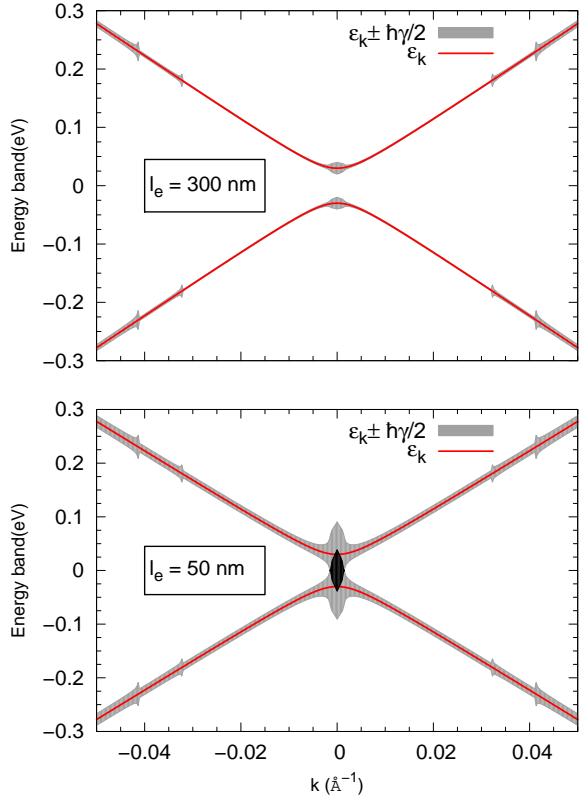}
%\vspace*{-0.4cm}
\caption{(Color online) 
Broadening of the electronic bands of the Qm tube
for different values of the elastic scattering lengths $l_e$.
The thin (red) solid lines are the energy bands $\epsilon_k$.
The gray area is the region delimited by the curves
$\epsilon_k\pm\hbar\gamma_{tot}(k)/2$, where $\gamma_{tot}(k)$ is the electronic
scattering rate.
For $l_e=50$~nm the grey areas of the two electronic bands are
superimposed in a region highlighted in black.}
\label{fig3}
\end{center}
\end{figure}

This counterintuitive behavior can be understood by considering 
the electronic scattering rate (or broadening) $\gamma_{tot}(k)$,
defined in Eq.~\ref{gamt}.
$\gamma_{tot}(k)$ represents the uncertainty with which the energy
of an electronic band can be defined.
Fig.~\ref{fig3} reports the electronic bands for the Qm tube:
the bands are represented as lines with a finite thickness
(gray area in Fig.~\ref{fig3}) equal to $\gamma_{tot}(k)$.
The increased value of $\gamma_{tot}(k)$ in the vicinity of the gap
is due to the elastic scattering.
The broadened electronic bands of Fig.~\ref{fig3}
overlap near $k=0$ for $l_e=50$~nm (low quality samples),
meaning that, because of energy-uncertainty, the electrons
do not see the gap and can tunnel from a band to the other even if the
applied field is small.
On the contrary, for $l_e=300$~nm (better quality samples) the bands do not overlap
and for small electric field the Zener tunneling can not take place.
In Fig.~\ref{fig3}, the sudden increase of the broadening for
$|k|>0.03$~\AA$^{-1}$ is due to the scattering with optical phonons.
This scattering mechanism is active only when
$|\epsilon(k)|>\Delta/2+\hbar \omega^{\bf K}$ and
$|\epsilon(k)|>\Delta/2+\hbar \omega^{\mathbf \Gamma}$,
which is far from the gap. Zener tunneling is, thus,
activated only by the elastic scattering (with defects) and not by the
scattering with optical phonons.

\begin{figure}
\begin{center} %\vspace*{-0.1cm}
\includegraphics[width = 0.44\textwidth]{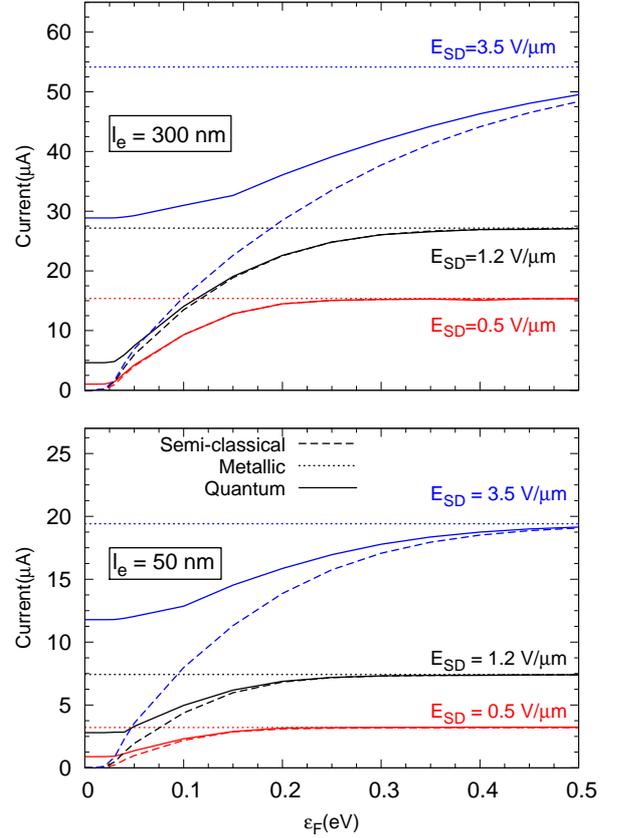}
%\vspace*{-0.52cm}
\caption{(Color online) Current versus $\epsilon _F$ for
different values of the electric field with $l_e=300$~nm and
$l_e=50$~nm.
Calculations for the Qm tubes are done with the semi-classical
or with the quantum approach. Calculations for metallic tubes
are also shown.
Note that, for $\epsilon _F \rightarrow 0$,
the current vanishes in the semi-classical approach,
while it has a finite value in the quantum approach.}
\label{cur5}
\end{center}
\end{figure}

Finally, to see the things from a different perspective, 
Fig. \ref{cur5}  reports the current as a function of $\epsilon_F$
for various values of the electric field $E_{SD}$.
For large doping (large $\epsilon_F$), the
Qm tube current calculated with the semiclassical and with the quantum
approaches tend to be equal. Also, both currents tend to be equal to
the metallic tube one.
The semi-classical approach predicts a zero current in the gap and
underestimates 
the current in a range of $\epsilon_F$ that increases with $E_{SD}$.
In the region of the gap, the current is thus entirely due to Zener tunneling.

\subsection{Current as a function of the electric field} 
\label{sec_current}

\begin{figure}
\begin{center} %\vspace*{-0.2cm}
\includegraphics[width = 0.5\textwidth]{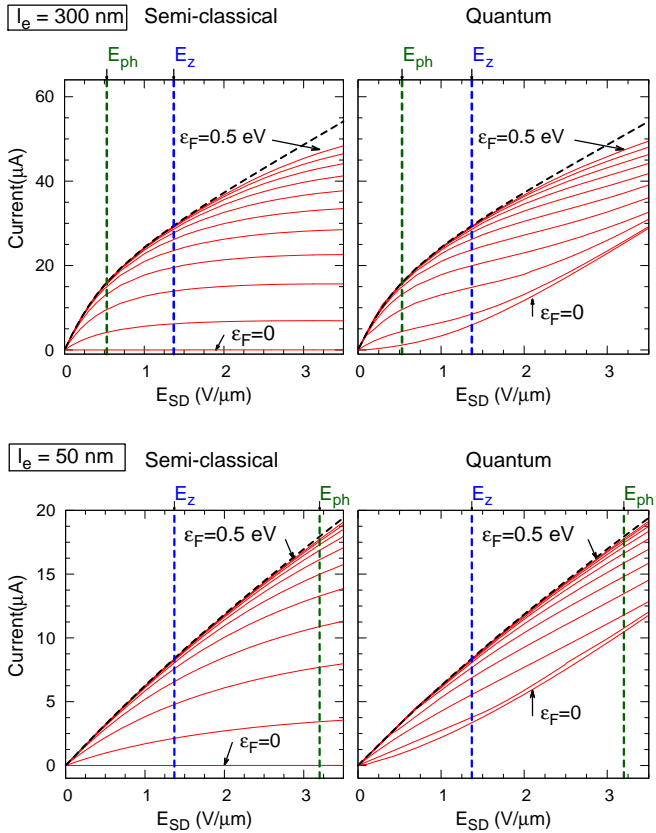}
%\vspace*{-0.52cm}
\caption{(Color online) Current versus. electric field $E_{SD}$ for
$l_e=300$~nm (top panels).  
$l_e=50$~nm (bottom panels).
Thin continuous lines (red): quasi metallic tube by varying $\epsilon_F$
from $0$ to $0.5$~eV with $0.05$~eV step.
Thick dashed lines: metallic tube.
Calculations are done with the semi-classical (left panels) or
the quantum (right panels) approaches.
$E_z$ and $E_{ph}$ are  
respectively the field from which Zener tunneling and the optical phonon
scattering become important.
$E_z \simeq 1.37 \mathrm{V/\mu m}$,
$E_{ph} \simeq 0.53$ $\mathrm{V/\mu m}$ for $l_e=300$~nm,
$E_{ph} \simeq 3.2$ $\mathrm{V/\mu m}$ for $l_e=50$~nm.
Note that for $\epsilon_F = 0$, the
current in the semi-classical model is exactly zero.}
\label{mcur5}
\end{center}
\end{figure}

To analyze the results as a function of the electric field $E_{SD}$, 
we consider two characteristic fields: $E_z$ and $E_{ph}$.
$E_z$ is defined in Eq.~(\ref{Ezener}) and it is the value of
the electric field for which, in the absence of scattering, the Zener
scattering is expected to be sizable.
$E_{ph}$ is defined as $E_{ph}=\hbar\omega^{\bf K}/(el_e)$ and it is
the value of the field for which optical-phonon emission become sizable
compared to elastic scattering.
$E_{ph}$ is higher for smaller values of $l_e$.
To understand this definition, one should consider that
optical phonons have a finite energy $\hbar\omega^\nu$.
In order to emit an optical phonon an electronic carrier needs to
attain a minimum activation energy corresponding to $\hbar\omega^\nu$.
This is possible if the carrier has accelerated in the electric
field $E_{SD}$ at least through a length $l_{act}=\hbar\omega
^\nu/(eE_{SD})$~\cite{kane}. The process is dominated by the zone boundary phonons
since $\omega^{\bf K}<\omega ^{\mathbf \Gamma}$.
For small electric fields, $E_{SD}\ll E_{ph}$, $l_e\ll l_{act}$ and
the scattering with defects prevents the carriers to accelerate and, thus,
to reach the activation energy.
On the other hand, if $E_{SD} \gg E_{ph}$, $l_e \gg l_{act}$ and the 
carriers can attain the activation energy before scattering with the defects.
Thus, for small electric field, elastic scattering is the dominant
scattering mechanism
(in this regime the current-voltage curve of metallic tubes is linear~\cite{kane}).
For electric fields higher that $E_{ph}$ the scattering with optical
phonons is activated.
This is associated with a decrease of the differential conductivity
(sub-linear current-voltage curve)
which can lead to a saturation of the current at high drain-source voltages
in metallic tubes with long elastic scattering lengths~\cite{kane}.
The scattering with optical phonons can lead to an anomalous
increase of the optical-phonon population (hot-phonons)~\cite{Lazzeri1,Lazzeri2}
which, in turn, will further diminish the high-bias conductivity.

Fig. \ref{mcur5}  shows the current as a function of the electric field for different values of
$\epsilon_F$.
At high doping, the results for 
the semi-classical (BTE) and of the  quantum (ME) approaches
are very similar and approach the metallic results.
The two approaches provide significantly different results for sufficiently high electric
fields and small values of the Fermi level.
In particular, for $\epsilon_F = 0$ the current is exactly zero within
the semi-classical model, while the current has a finite
value, which increases with the electric field, within the quantum
approach, Fig.~\ref{mcur5}.
For $\epsilon_F =0$ and $\epsilon_F=0.05$~eV, within the quantum
approach the current versus. electric field is a super-linear curve for $E_{SD}\sim E_z$. 
Such a behavior was observed in graphene \cite{Niels} and in large diameter nanotubes \cite{Anantram}, and identified as due to Zener tunneling.

Let us compare the results for the different scattering lengths $l_e$ (Fig.~\ref{mcur5}).
Zener tunneling should be more visible for $l_e=$ 50~nm, than for $l_e=$ 300~nm,
for two reasons.
First, for $l_e=$ 50~nm, the Zener current is already relevant for electric
fields smaller than $E_z$ (this is explained in the discussion related to Fig.~\ref{fig3}).
Second, for $l_e=$ 50~nm $E_{ph} > E_z$, while for $l_e=$ 300~nm $E_{ph}<E_z$.
Now, the scattering with optical phonons is associated to a sub-linear
current which tend to mask the presence of the super-linear Zener
current. 
For $l_e=$ 50~nm, $E_{ph}>E_z$ means that this ``masking'' does not occur in the
region of interest.
For $l_e=$ 300~nm, $E_{ph}<E_z$ means that this ``masking'' is already active in the
region where Zener is active.
Beside this, one should also consider that in the present simulations
we are not including the possibility of hot phonons. The
inclusion of this effect would decrease the differential conductance for
$E_{SD}>E_{ph}$.
This should not change the $l_e=$ 50~nm curves but should further mask the
Zener current in the $l_e=$ 300~nm case.

\begin{figure}[h!]
\begin{center} %\vspace*{-0.cm}
\includegraphics[width = 0.48\textwidth]{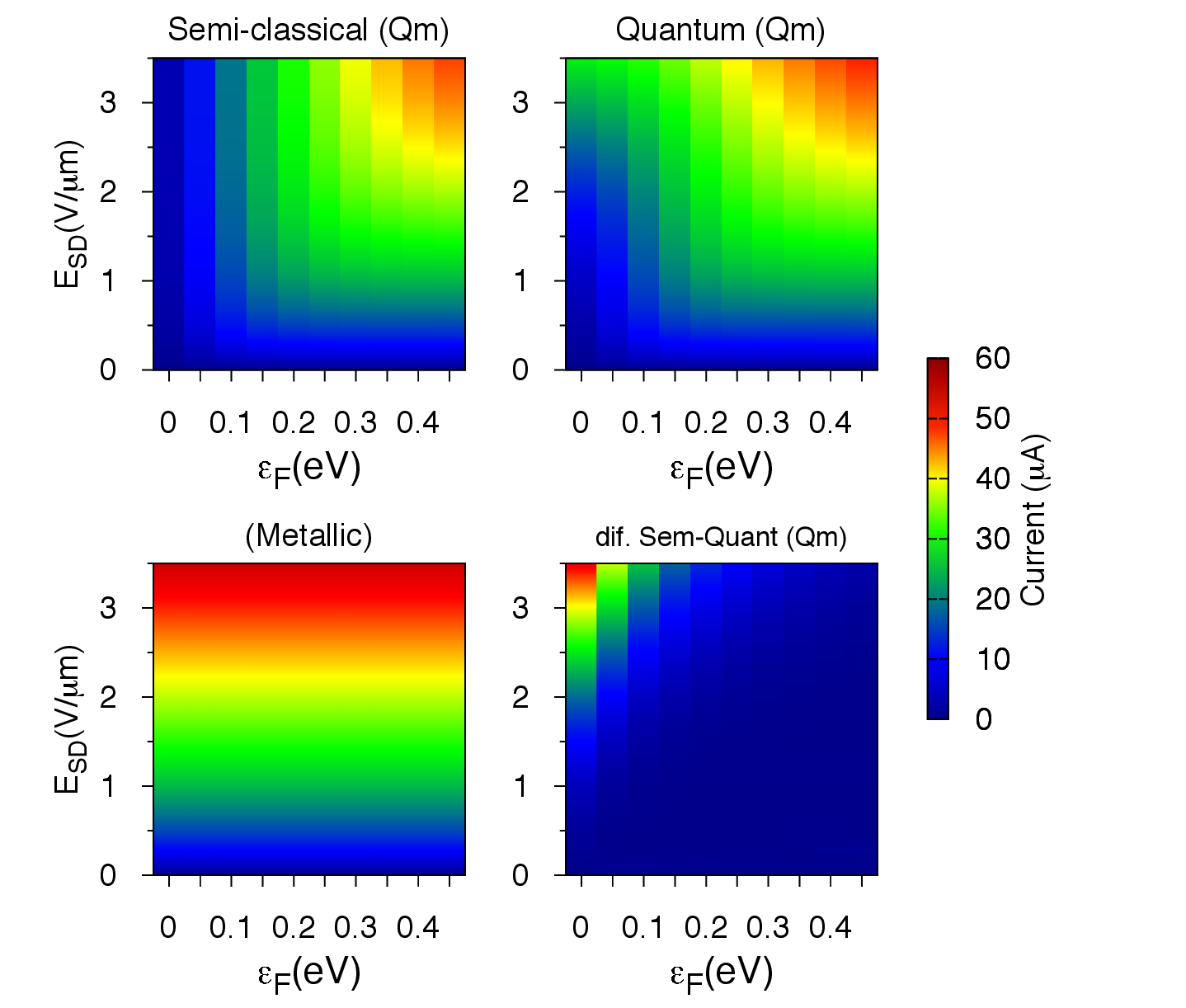}
%\vspace*{-0.3cm}
\caption{(Color online) Current as a function of the electric field $E_{SD} $
and of the Fermi level $\epsilon_F$ for an elastic scattering
$l_e=300$~nm.
For the quasi-metallic tube (Qm), calculations are done
using the semi-classical or the quantum approaches.
The difference between the semi-classical
and quantum calculations [dif. Sem-Quant (Qm)] is also shown and the
current for this panel is multiplied by a factor 2.  Calculations for
the metallic tube are shown as a comparison.}
\label{map3}
\end{center}
%\end{figure}

%\begin{figure}
 \vspace{-0.4cm}
\begin{center}
\includegraphics[width =0.48\textwidth]{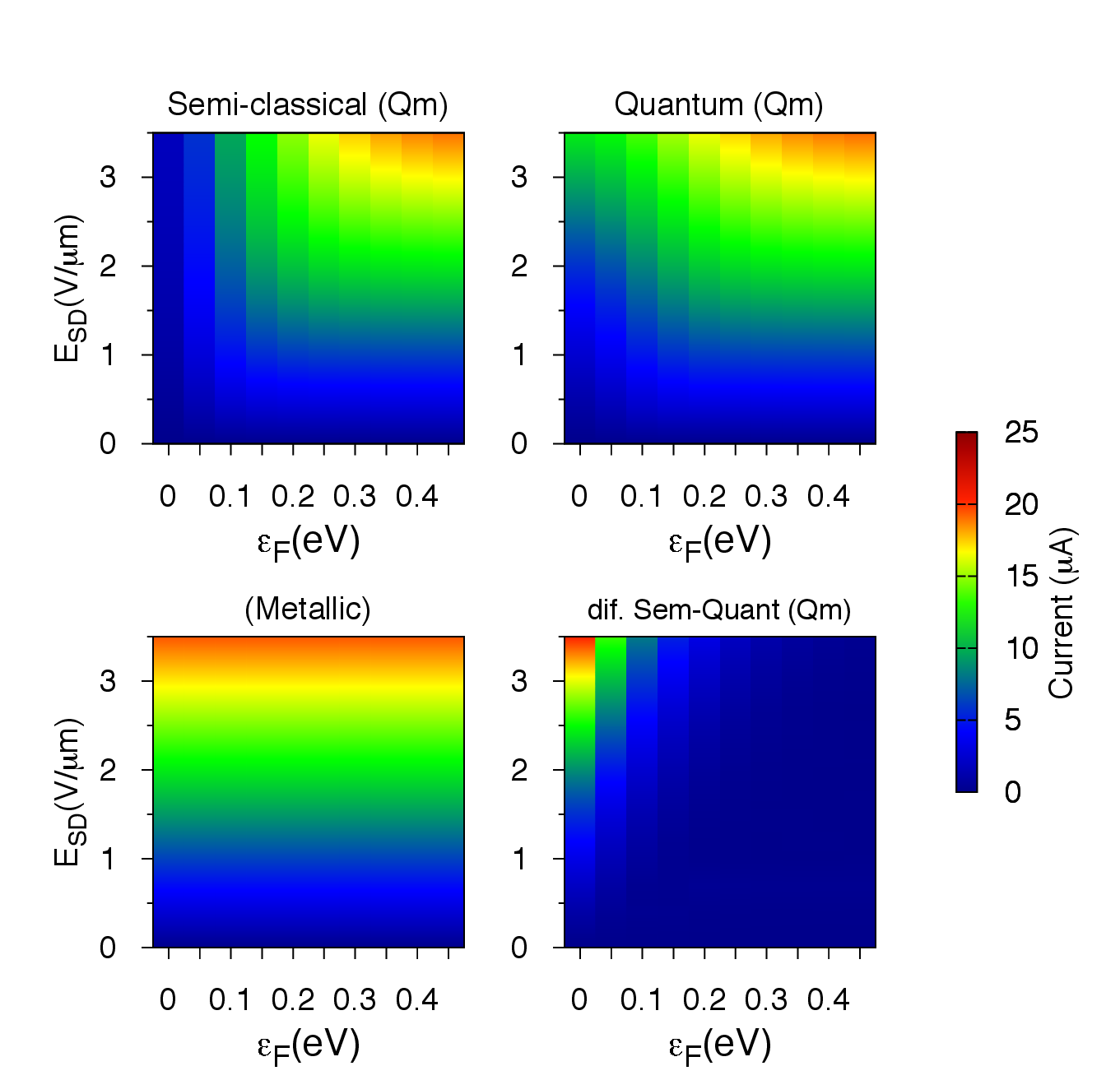}
%\vspace*{-0.3cm}
\caption{(Color online)
Current as a function of $E_{SD} $ and $\epsilon_F$ for an elastic scattering $l_e=50$~nm.
Same caption as in Fig.~\ref{map5}.}
\label{map5}
\end{center}
\end{figure}

Finally, in by Fig.\ref{map3} and \ref{map5} we present the two-dimensional maps
of the current as a function of $E_{SD}$ and $\epsilon_F$.
In one of the panels, we also report the difference between the
current obtained within the quantum approach and that of the semi-classical
calculation. This difference is higher where the Zener tunneling is
more important.
For metallic tubes, the current does not depend on the
Fermi level because, within the linear band approximation, the
electronic density of states is a constant. Note that for sufficiently
high $\epsilon_F$ values the semi-classical and the quantum
results for the Qm tubes are almost indistinguishable and approach the 
results for the metallic tubes.

\subsection{Steady-state distribution functions} 

\begin{figure}
\begin{center} %\vspace*{-2.2cm}
\includegraphics[width = 0.48\textwidth]{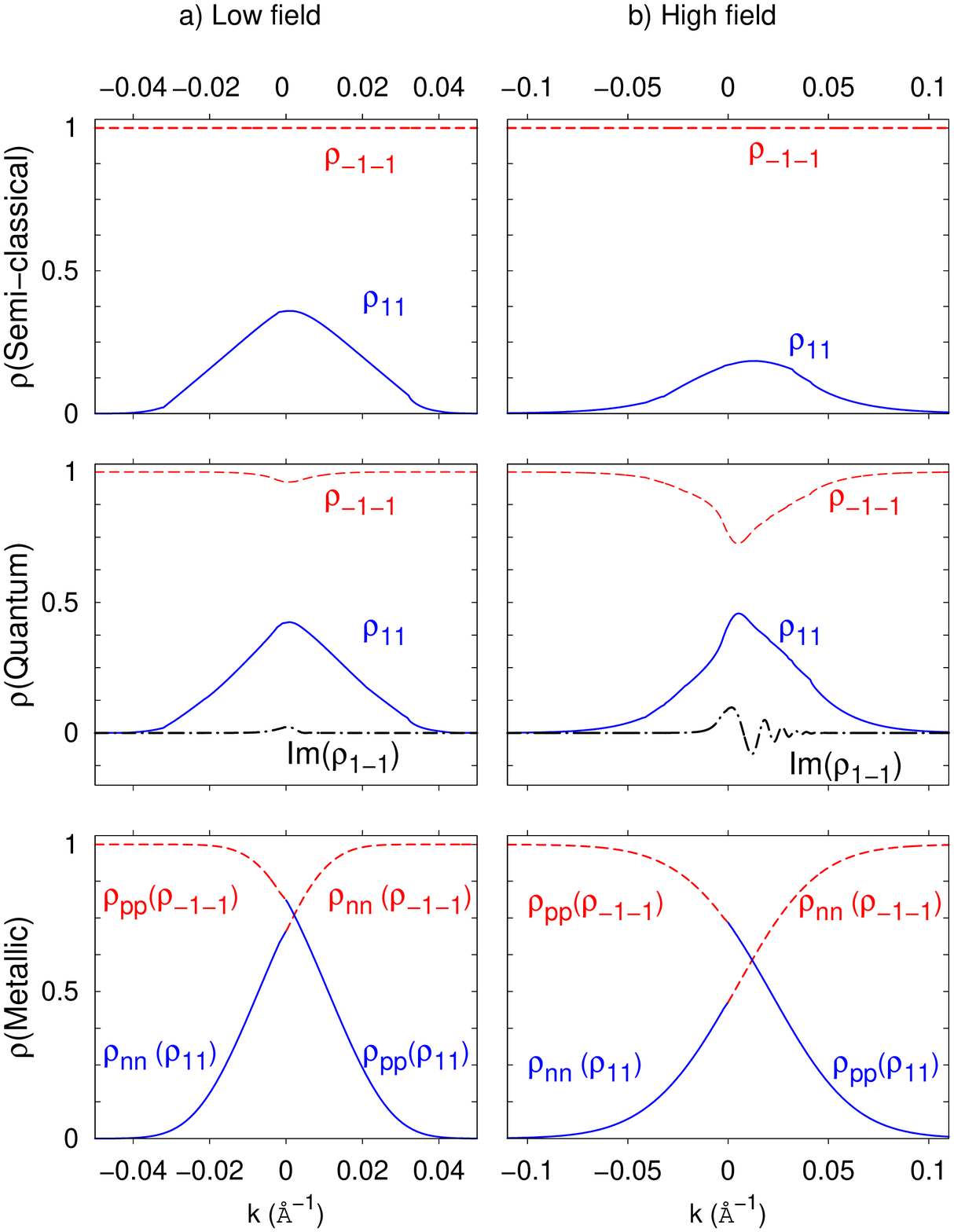}
%\vspace*{-0.3cm}
\caption{(Color online) 
Equilibrium distribution functions for conduction ($\rho_{11}$) and
valence bands ($\rho_{-1-1}$) and imaginary part of the off-diagonal
($\mathrm{Im}(\rho_{1-1})$) of the density matrix.
Calculations are done with $l_e=50 \mathrm{nm}$ and $\epsilon_F = 0.05$~eV,
for two different electric fields $E_{SD}$.
The three panels in the left column correspond to $E_{SD} =
0.5$~V/$\mu$m (low filed),
while those on the right column to $E_{SD} = 3.5$~V/$\mu$m (high field).
The two panels of the topmost row are calculations done for Qm tubes
within the semi-classical (BTE) approach. The middle row panels are Qm tubes done
within the quantum approach (ME). The bottom row panels correspond to
metallic tubes. In this case $\rho_{nn}$ ($\rho_{pp}$) is the distribution function of the band with negative (positive) velocity
(see Fig.~\ref{qmb}).}
\label{disthlf}
\end{center}
\end{figure}

It is instructive to compare the steady-state distribution functions
of the electronic populations associated with the semi-classical
or quantum approaches.
Fig.~\ref{disthlf} reports the distribution functions obtained for 
$l_e=50$~nm and the Fermi level $\epsilon_F = 0.05$~eV,
that is for conditions in which the quantum effects are relevant.
In Qm tubes, when the quantum approach is used both electron and hole 
contribute to the current, whereas in the semiclassical approach
only electrons carry a current 
since the valence band is filled. 
Interestingly, at high field the population of Qm tubes tends to that
of metallic tubes only when the quantum approach is used. Indeed, in the metallic tube (which corresponds to the limit $\Delta  \rightarrow 0$) 
the tunneling  between valence and conduction band is total. The metallic limit can be reached only when one includes interband transitions.
Fig.~\ref{disthlf} also shows the imaginary part of the off-diagonal element of the density matrix
$\mathrm{Im}(\rho_{1-1})$. This quantity is different from zero when
quantum effects are relevant. One can remark that Zener tunneling
turns on near $k = 0$ (with a maximum at $k = 0$ ) and decays after
few oscillations for $k>0$ . The typical wave vector associated with these oscillations
is $\propto 1/\Delta$, $\Delta$ being the electronic gap.

\section{Conclusion}

We have presented a theoretical model based on the master equation
to explore the quantum effects (inter-band transitions induced
by electric field or Zener tunneling) on the transport properties of
homogeneous quasi-metallic carbon nanotubes (Qm).
The presence and relevance of the Zener tunneling has been highlighted
by comparing with a semi-classical Boltzmann transport approach,
in which the inter-band transitions are neglected.
We studied Qm tube with an electronic gap $\Delta=50$~meV
in the presence of impurities or defects which influence the transport by providing
an elastic scattering channel for the carriers.
Zener tunneling is relevant for small doping, when the Fermi energy lies in or close to the 
forbidden gap $\Delta$.
In absence of elastic scattering (in high quality samples), the small size of the band gap of Qm tubes enables Zener tunnelling for realistic values of the the electric field (above $\sim$ 1 V/$\mu$m).
However, for such electric fields the scattering with the
optical phonons (whose effect is underestimated in the present approach)
tends to mask the Zener current.
On the other hand, the presence of a strong elastic scattering (in low quality samples) further decreases the values of the field required to observe Zener tunnelling.
Indeed, for elastic-scattering lengths of the order of 50 nm, Zener tunnelling affects the current/voltage characteristic already in the linear regime.
In other words, Zener tunneling is made visible by defects in Qm
tubes. This result is similar to what has been recently shown for graphene \cite{Niels}.

Calculations were done at IDRIS, Orsay, project 096128.

%\twocolumn

%\bibliography{biblio}
%\end{multicols}
\end{document}